\documentstyle[epsfig]{aipproc}

\begin{document}
\title{Threshold Resummation and the Total Cross Section for Top
Quark Production\footnote
{Paper presented by E. L. Berger, to be published in the Proceedings 
of DIS'97, Fifth International Workshop on Deep Inelastic Scattering and QCD, 
Chicago, IL, April 14 - 18, 1997.  Argonne report ANL-HEP-CP-97-33.  This work 
was supported by the US Department of Energy, Division of High Energy Physics, 
Contract No.W-31-109-ENG-38.}}

\author{Edmond L. Berger and Harry Contopanagos}
\address{High Energy Physics Division\\
Argonne National Laboratory, Argonne, Illinois 60439}

\maketitle
\vspace{-10pt}
\begin{abstract}
We discuss the motivation for resummation of the effects of initial-state
soft gluon radiation, to all orders in the strong coupling strength, for 
processes in which the near-threshold region in the partonic subenergy is 
important.  We summarize our calculation of the total cross section for top 
quark production at hadron colliders.  Comments are included on the differences 
between our treatment of subleading logarithmic terms and other methods.
\end{abstract}
\vspace{-10pt}
\section*{Introduction and Motivation}

In inclusive hadron interactions at collider energies, 
$t\bar{t}$ pair production proceeds through partonic hard-scattering 
processes involving initial-state light quarks $q$ and gluons $g$.  In 
lowest-order perturbative quantum chromodynamics (QCD), at 
${\cal O}(\alpha_s^2)$,  the two partonic subprocesses 
are $q + \bar{q} \rightarrow t + \bar{t}$ and $g + g \rightarrow t + \bar{t}$.  
Calculations of the cross section through next-to-leading order, 
${\cal O}(\alpha_s^3)$, involve gluonic radiative corrections to these 
lowest-order subprocesses as well as contributions from the $q + g$ initial 
state~\cite{dawson}.  In this paper, we describe calculations that go 
beyond fixed-order perturbation theory through resummation of the effects of 
gluon radiation~\cite{laeneno,previous,catani} to all orders in 
the strong coupling strength $\alpha_s$.  

The physical cross section is obtained through the factorization theorem
\begin{equation}
\sigma_{ij}(S,m)=
{4m^2\over S}\int_0^{{S\over 4m^2}-1}d\eta
\Phi_{ij}(\eta,\mu) \hat\sigma_{ij}(\eta,m,\mu) .
\label{feleven}
\end{equation}
%
The square of the total hadronic center-of-mass energy is $S$, the square of 
the partonic center-of-mass energy is $s$, $m$ denotes the top mass, $\mu$ is 
the usual factorization and renormalization scale, and $\Phi_{ij}(\eta,\mu)$ is 
the parton flux.  
The variable $\eta={s \over 4m^2} - 1$ measures the distance from the 
partonic threshold.  The indices $ij\in\{q\bar{q},gg\}$ denote the initial 
parton channel.  The partonic cross section 
$\hat\sigma_{ij}(\eta,m,\mu)$ is obtained either from fixed-order QCD
calculations~\cite{dawson}, or, as described here, from calculations
that include of resummation~\cite{laeneno,previous,catani} to all 
orders in $\alpha_s$.  We use the notation 
$\alpha \equiv \alpha(\mu=m) \equiv \alpha_s(m)/\pi$.  The total physical 
cross section is obtained after incoherent addition of the contributions from 
the the $q\bar{q}$ and $gg$ production channels.  

Comparison of the partonic cross section at next-to-leading order with its 
lowest-order value reveals that the ratio becomes very large in the 
near-threshold region.  Indeed, as $\eta \rightarrow 0$, the ``$K$-factor" at 
the partonic level $\hat K(\eta)$ grows in proportion to $\alpha \ln^2(\eta)$. 
The very large mass of the top quark notwithstanding, the large ratio 
$\hat K(\eta)$ makes it evident that the next-to-leading order result does not 
necessarily provide a reliable quantitative prediction of the top quark 
production cross section at the energy of the Tevatron collider.  Analogous 
examples include the production of hadronic jets that carry large values of 
transverse momentum and the production of pairs of supersymmetric particles 
with large mass.  

\section*{Gluon Radiation and Resummation}

The origin of the large threshold enhancement may be traced to initial-state
gluonic radiative corrections to the lowest-order channels.  
We remark that we are calculating the inclusive total cross 
section for the production of a top quark-antiquark pair, i.e., the total 
cross section for $t + \bar{t} + \rm anything$.  The partonic subenergy 
threshold in question is the threshold for $t + \bar{t} +$ any number of 
gluons.  This coincides with the threshold in the invariant mass of the 
$t + \bar{t}$ system for the lowest order subprocesses only. 

For $i + j \rightarrow t + \bar{t} + g$, we define the variable $z$ through 
the invariant
$(1-z) = {2k \cdot p_t \over m^2}$, where $k$ and $p_t$ are the four-vector 
momenta of the gluon and top quark.  In the limit that 
$z \rightarrow 1$, the radiated gluon carries zero momentum.  After cancellation
of soft singularities and factorization of collinear singularities in 
${\cal O}(\alpha_s^3)$, there is a left-over integrable large logarithmic 
contribution to the partonic cross section associated with initial-state gluon 
radiation.  This contribution is often expressed in terms of ``plus" 
distributions.  In ${\cal O}(\alpha_s^3)$, it is proportional to 
$\alpha^3 \ln^2(1-z)$.  When integrated over the near-threshold 
region $1 \ge z \ge 0$, it provides an excellent approximation to the full 
next-to-leading order physical cross section as a function of the top mass.  
The goal of gluon resummation is to sum the series in 
$\alpha^{n+2} \ln^{2n}(1-z)$ to all orders in $\alpha$ in order to 
obtain a more trustworthy prediction.

Different methods of resummation differ in theoretically and phenomenologically
important respects.  Formally, if not explicitly in some approaches, an 
integral over the radiated gluon momentum $z$ must be done over regions in 
which $z \rightarrow 1$.  Therefore, one significant distinction among methods 
has to do with how the inevitable ``non-perturbative" region is handled.  

The method of resummation we employ~\cite{previous} is based on a 
perturbative truncation of principal-value (PV) 
resummation~\cite{stermano}.  
This approach has an important technical advantage in that it 
does not depend on arbitrary infrared cutoffs.  Because extra scales are 
absent, the method permits an evaluation of its perturbative regime of
applicability, i.e., the region of the gluon radiation phase
space where perturbation theory should be valid.  We work in the 
$\overline{\mbox{MS}}$ factorization scheme.

Factorization and evolution lead directly to exponentiation of the set of 
large threshold logarithms in moment ($n$) space in terms of an exponent 
$E^{PV}$.  The function $E^{PV}$ is finite, and 
$\lim_{n\rightarrow\infty}E^{PV}(n,m^2)=-\infty$.  Therefore, 
the corresponding partonic cross section is finite as $z\rightarrow 1
\ (n\rightarrow +\infty)$.
The function $E^{PV}$ includes both perturbative and non-perturbative content.  
The non-perturbative content is not a prediction of perturbative QCD. 
We choose to use the exponent only in the interval in moment space in which the 
perturbative content dominates.  We derive a 
perturbative asymptotic representation of $E(x,\alpha(m))$ that is 
valid in the moment-space interval
\begin{equation}
1<x\equiv \ln n< t\equiv {1\over 2\alpha b_2}.
\label{tseven}
\end{equation}
The coefficient $b_2=(11C_A-2n_f)/12$; the number of flavors $n_f=5$; 
$C_{q\bar{q}}=C_F=4/3$; and $C_{gg}=C_A=3$. 

The perturbative asymptotic representation is
\begin{equation}
E_{ij}(x,\alpha)\simeq E_{ij}(x,\alpha,N(t))=
2C_{ij}\sum_{\rho=1}^{N(t)+1}\alpha^\rho
\sum_{j=0}^{\rho+1}s_{j,\rho}x^j\ .
\label{teight}
\end{equation}
Here
\begin{equation}
s_{j,\rho}=-b_2^{\rho-1}(-1)^{\rho+j}2^\rho c_{\rho+1-j}(\rho-1)!/j!\ ;
\label{tnine}
\end{equation}
and $\Gamma(1+z)=\sum_{k=0}^\infty c_k z^k$, where $\Gamma$ is the Euler gamma 
function.  
The number of perturbative terms $N(t)$ in Eq.~(\ref{teight}) is
obtained~\cite{previous} by optimizing the asymptotic approximation
$\bigg|E(x,\alpha)-E(x,\alpha,N(t))\bigg|={\rm minimum}$. 
Optimization works perfectly, with $N(t)=6$ at $m = 175$ GeV.   
As long as $n$ is in the interval of Eq.~(\ref{tseven}),
all the members of the family in $n$ are optimized 
at the same $N(t)$, showing that the optimum number of 
perturbative terms is a function of $t$, i.e., of $m$ only.

Resummation is completed in a finite number of steps.  Upon using the running 
of the coupling strength $\alpha$ up to two loops only, 
all monomials of the form $\alpha^k\ln^{k+1}n,\ \alpha^k\ln^kn$
are produced in the exponent of Eq.~(\ref{teight}).  We discard monomials
$\alpha^k\ln^kn$ in the exponent because of the restricted leading-logarithm
universality between $t\bar{t}$ production and massive lepton-pair 
production, the Drell-Yan process.

The exponent we use is the truncation
\begin{equation}
E_{ij}(x,\alpha,N)=2C_{ij}\sum_{\rho=1}^{N(t)+1}\alpha^\rho s_\rho x^{\rho+1} ,
\label{tseventeen}
\end{equation}
with the coefficients
$s_\rho\equiv s_{\rho+1,\rho}=b_2^{\rho-1}2^\rho/\rho(\rho+1)$.  This
expression contains no factorially-growing (renormalon) terms. 
It is valuable to stress that we can derive the perturbative expressions,
Eqs.~(\ref{tseven}), (\ref{teight}), and (\ref{tnine}),  without the 
principal-value prescription, although with less certitude~\cite{previous}.

After inversion of the Mellin transform from moment space to 
the physically relevant momentum space, the resummed partonic cross sections, 
including all large threshold corrections, can be written
\begin{equation}
\hat{\sigma}_{ij}^{R;pert}(\eta,m)=
\int_{z_{min}}^{z_{max}}dz
{\rm e}^{E_{ij}(\ln({1\over 1-z}),\alpha)}
\hat{\sigma}_{ij}'(\eta,m,z) .
\label{bthreep}
\end{equation}
The leading large threshold corrections are contained in the exponent 
$E_{ij}(x,\alpha)$, a calculable polynomial in $x$.  The derivative
$\hat{\sigma}_{ij}'(\eta,m,z)=d(\hat{\sigma}_{ij}^{(0)}(\eta,m,z))/dz$,
and $\hat{\sigma}_{ij}^{(0)}$ is the lowest-order ${\cal O}(\alpha_s^2)$
partonic cross section expressed in terms of inelastic kinematic variables.
The lower limit of integration, $z_{min}$, is fixed by kinematics.  The upper 
limit, $z_{max} < 1$, is set by the boundary between the perturbative and 
non-perturbative regimes, well specified within the context of our 
calculation.  Its presence assures us that our integration over the soft-gluon 
momentum is carried out only over a range in which poorly specified 
non-universal subleading terms would not contribute significantly even if 
retained.  We cannot justify continuing the results of leading-logarithm 
resummation into the region $z > z_{max}$.

Perturbative resummation probes the threshold down to
$\eta\ge \eta_0 =(1-z_{max})/2 $.  Below this value, perturbation theory is 
not to be trusted.  For $m$ = 175 GeV, we determine that the perturbative 
regime is restricted to values of the subenergy greater than 1.22 GeV above the 
threshold ($2m$) in the $q{\bar q}$ channel and 8.64 GeV 
above threshold in the $gg$ channel. The difference reflects the larger
color factor in the $gg$ case.  The value 1.22 GeV is comparable to the decay 
width of the top quark, a natural definition of the perturbative boundary and 
by no means unphysically large.    

\section*{Physical cross section}

Other than the top mass, the only undetermined scales are the QCD 
factorization and renormalization scales.  We adopt a common value $\mu$ for 
both.
\begin{figure}[b!]
\centerline{\epsfig{file=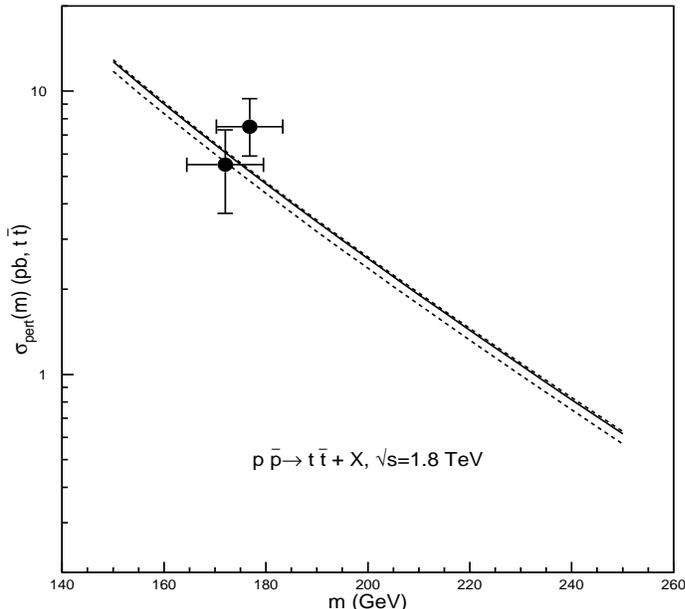,height=2.75in,width=2.75in}}
\vspace{10pt}
\caption{Inclusive total cross section for top quark production.  The dashed 
curves show the upper and lower limits while the solid curve 
is our central prediction.  CDF and D0 data are shown.}
\label{fig1}
\end{figure}
In Fig.~1, we show our total cross section for $t\bar{t}$-production 
as a function of top mass in $p \bar{p}$ collisions 
at $\sqrt{S}=1.8$ TeV.  The central value is obtained with the 
choice $\mu/m=1$, and the lower and upper limits are  the maximum 
and minimum of the cross section in the range $\mu/m\in\{0.5,2\}$.  
At $m =$ 175 GeV, the full width of the uncertainty band 
is about 10\%\ .  As is to be expected, less variation with $\mu$ is evident 
in the resummed cross section than in the next-to-leading order cross section.  
In estimating uncertainties, we do not consider explicit variations of our 
non-perturbative boundary, expressed through $z_{max}$.  
This is justified because, for a fixed $m$ and $\mu$, $z_{max}$
is obtained by enforcing dominance of the universal leading logarithmic terms 
over the subleading ones. Therefore, $z_{max}$
is {\it derived} and is not a source of uncertainty.  At fixed $m$, the 
boundary necessarily varies as $\mu$ and thus $\alpha$ vary. 

Our calculation is in agreement with the  data~\cite{cdfdz}.  We find
$\sigma^{t\bar{t}}(m=175\ {\rm GeV},\sqrt{S}=1.8\ {\rm TeV})=
5.52^{+0.07}_{-0.42}\ \rm{pb}$.  This cross section is larger than the 
next-to-leading order value by about $9\%$.  The top quark cross section 
increases quickly with the energy of the $p \bar{p}$ collider.  We determine 
$\sigma^{t\bar{t}}(m=175\ {\rm GeV},\sqrt{S}=2\ {\rm TeV})=
7.56^{+0.10}_{-0.55}\ \rm {pb}$.  The central value rises to 22.4 pb at 
$\sqrt{S}=3\ {\rm TeV}$ and 46 pb at $\sqrt{S}=4\ {\rm TeV}$.

Extending our calculation to much larger values of 
$m$ at ${\sqrt S}=1.8$ TeV, we find that resummation in the principal 
$q\bar{q}$ channel produces 
enhancements over the next-to-leading order cross section of $21\%$, $26\%$, 
and $34\%$, respectively, for $m =$ 500, 600, and 700 GeV.  The reason for the
increase of the enhancements with mass at fixed energy is that the threshold 
region becomes increasingly dominant.  Since the $q\bar{q}$ 
channel also dominates in the production of hadronic jets at very large values 
of transverse momenta, we suggest that on the order of $20\%$ of the excess
cross section reported by the CDF collaboration~\cite{cdfjets} may be 
accounted for by resummation.

\section*{Other Methods of Resummation}
Two other groups have published calculations of the total cross section at 
$m=175\ {\rm GeV}$ and $\sqrt{s}=1.8\ {\rm TeV}$:
$\sigma^{t\bar t}$({\rm LSvN}~\cite{laeneno}) = $4.95^{+0.70}_{-0.40}$ pb; 
and  
$\sigma^{t\bar t}$({\rm CMNT}~\cite{catani}) = $4.75^{+0.63}_{-0.68}$ pb.  
From a numerical point of view, all agree 
within their estimates of theoretical uncertainty.  However, the resummation 
methods differ as do the methods for estimating uncertainties.  Both 
the central value and the band of uncertainty of the LSvN predictions are 
sensitive to their infrared cutoffs.  To estimate theoretical 
uncertainty, we use the standard $\mu$-variation, whereas LSvN obtain theirs 
primarily from variations of their cutoffs.  It is difficult to be certain 
of the central value and to evaluate theoretical uncertainties in a method 
that requires an undetermined infrared cutoff.  

The group of Catani, Mangano, Nason, and Trentadue (CMNT)~\cite{catani} 
calculate a central value of 
the resummed cross section (also with $\mu/m = 1$) that is less than 
$1\%$ above the exact next-to-leading order value.  
There are similarities and differences between our approach 
and the method of CMNT.  We use the same universal leading-logarithm 
expression in moment space, but differences occur after the transformation to
momentum space.  The differences can 
be stated more explicitly if we examine the perturbative expansion of the
resummed hard kernel ${\cal H}^{R}_{ij}(z,\alpha)$.   
If, instead of 
restricting the resummation to the universal leading logarithms only, we were 
to use the full content of ${\cal H}^{R}_{ij}(z,\alpha)$, we would arrive at 
an analytic expression that is equivalent to the numerical inversion of 
Ref.~\cite{catani}, 
\begin{equation}
{\cal H}^{R}_{ij} \simeq 1+2\alpha C_{ij} 
\biggl[\ln^2 (1-z) + 2\gamma_E \ln (1-z)\biggr] + {\cal O}(\alpha^2).
\label{padovao}
\end{equation} 
In terms of this expansion, in our work we retain only the leading term 
$\ln^2 (1-z)$ at order $\alpha$, but both this term and the non-universal 
subleading term $ 2\gamma_E \ln (1-z)$ are retained in Ref.~\cite{catani}. 
If this subleading term is discarded in Eq.~(\ref{padovao}), the
residuals $\delta_{ij}/\sigma_{ij}^{NLO}$ defined in Ref.~\cite{catani} 
increase from 
$0.18\%$ to $1.3\%$ in the $q\bar{q}$ production channel and from $5.4\%$ to 
$20.2\%$ in the $gg$ channel.  After addition of the two 
channels, the total residual $\delta/\sigma^{NLO}$ grows from the negligible 
value of about $0.8\%$ to the value $3.5\%$.  While still smaller than 
the increase of $9\%$ that we obtain, the increase of $3.5\%$ vs. $0.8\%$ 
shows the substantial influence of the subleading logarithmic terms retained
in Ref.~\cite{catani}. 

We judge that it is preferable to integrate over only the region of phase 
space in which the subleading term is suppressed numerically.  Our reasons 
include the fact that the subleading term is not universal, is not the same 
as the subleading term in the exact ${\cal O}(\alpha^3)$ calculation and 
can be changed if one elects to keep non-leading terms in moment space.  The 
subleading term is negative and numerically very significant when it is  
integrated throughout phase space (i.e., into the region of $z$ above our 
$z_{max}$).  In the $q\bar{q}$ channel at $m=175$ GeV and ${\sqrt S}=1.8$ TeV, 
its inclusion eliminates more than half of the contribution from the leading 
term.  Although the goal is to resum the threshold contributions responsible 
for the enhancement of the cross section at next-to-leading order, the method 
of Ref.~\cite{catani} does not reproduce most of this enhancement.  The 
influence of non-universal subleading terms is amplified at higher orders 
where additional subleading structures occur in the approach of 
Ref.~\cite{catani} with significant numerical coefficients proportional 
to $\pi^2$, $\zeta(3)$, and so forth.  In our view, the results of a 
{\it leading-logarithm} resummation should not rely on subleading structures 
in any significant manner.  The essence of our determination of the 
perturbative boundary $z_{max}$ is precisely that below $z_{max}$ subleading 
structures are also numerically subleading, whether or not classes of poorly 
substantiated subleading logarithms are included.  For a more detailed 
discussion of these and other points of difference, including issues of 
momentum conservation and factorial growth, see section VI of 
Ref.~\cite{latest}.

Since the large threshold logarithms are mastered by resummation, the 
theoretical reliability of the resummed result is greater than 
that of a fixed-oder calculation.  Our analysis and the stability of our cross 
section under variation of the hard scale $\mu$ provide confidence that our 
perturbative resummation yields an accurate calculation of the inclusive top 
quark cross section at Tevatron energies.
%

\end{document}